\documentclass[12pt,preprint]{aastex}

\slugcomment{Submitted to PASP}

\shorttitle{Galactic--Scale Star Formation}
\shortauthors{Calzetti \& Kennicutt}

\begin{document}

\title{The New Frontier: Galactic--Scale Star Formation}

\author{D. Calzetti\altaffilmark{1},  R. C. Kennicutt\altaffilmark{2}} 

\altaffiltext{1}{Dept. of Astronomy, University of Massachusetts, Amherst, MA 01003; calzetti@astro.umass.edu}
\altaffiltext{2}{Institute of Astronomy, Cambridge University, Cambridge, U.K.}

\begin{abstract}
The arena of investigation of star formation and its scaling laws is slowly, but consistently, shifting from 
the realm of luminous galaxies to that of faint ones and to sub--galactic regions, as existing and new 
facilities enable investigators to probe regions of the combined parameter space of surface brightness, 
wavelength, and angular resolution that were inaccessible until a few years ago. We summarize what has been accomplished, and what remain as challenges in the field of galactic--scale star formation.
\end{abstract}

\keywords{Galaxies; ISM}

\section{Introduction}

Over the past two decades, evidence has been increasingly accumulating that there is a tight 
relation between the star formation rate surface density and the gas surface 
density {\em on global (disk--averaged) scales}  in nearby galaxies, which is expressed, using the 
parametrization of  \citet{Schmidt1959,Schmidt1963}, as:
\begin{equation}
\Sigma_{SFR} = A  \Sigma_{gas}^N, 
\end{equation}
where $\Sigma_{SFR}$ is in units of M$_{\odot}$~yr$^{-1}$~kpc$^{-2}$ and $\Sigma_{gas}$ in 
M$_{\odot}$~pc$^{-2}$, and with N$\approx$1.4--1.5 and A$\approx$2.5$\times$10$^{-4}$ 
\citep[e.g.,][and references therein]{Kennicutt1998a}. 

The scaling relation provides a direct link between the gas supply and the efficiency of the conversion 
process of gas into stars. Implicitly included in the physical mechanisms that regulate star formation 
is the threshold of star formation, i.e., the minimum gas surface density 
below which star formation  cannot be initiated \citep[e.g.,][]{Martin2001}. Understanding the physical 
connection between star formation and its 
fuel is a critical ingredient for models of the evolution of galaxies and their baryonic component 
\citep[e.g.,][]{Kay2002}.  A variety of physical models have been proposed to explain  the power--law scaling between gas and star formation surface density and the presence of a threshold for star formation, including large--scale gravitational instabilities \citep{Martin2001,Elmegreen2002,Wong2002}, local dynamical timescales of 
rotating disks  \citep{Wyse1989,Kennicutt1998a},  galactic shear  \citep{Hunter1998}, turbulence 
and cloud--cloud collision mechanisms or local gravity \citep{MacLow2004, Krumholz2005, Tasker2008, Heitsch2008, Krumholz2009}, and many others. These 
models cannot, however, be discriminated by global galaxy measures or by measurements that only  target the brightest, most intense star--forming, and densest environments within galaxies. 

The arena of investigation has been progressively shifting  from large (global) scales to small (sub--kpc) 
scales and from bright to faint regions and galaxies in an attempt, among other things, to break this 
degeneracy, and determine the physical underpinning of the scaling laws of star formation. 
Over the past several years, studies have expanded the investigation 
of the relation between star formation and gas to radial profiles of galaxies, to constrain the form of the relation over a 
few kpc scales \citep[e.g.,][and many others]{Martin2001, Boissier2003, Schuster2006}. Only with the relatively recent accomplishment  of high spatial resolution mapping in CO and HI to probe the resolved 
molecular and atomic gas content of galaxies,  and of homogeneous, arcsecond--resolution,    
UV--to--infrared multi--wavelength imaging surveys to 
derive robust dust--corrected SFRs, gas and star formation are beginning to the traced on the 
sub--kpc  scales typical of star--forming regions in galaxies \citep{Kennicutt2007, Bigiel2008, Leroy2008}. These studies have also yielded larger variations in the index of the power law, with values in the range N$\approx$1--3.  This result underscores the fact that challenges are still present not only on the theoretical front, but on the observational front as well. 

\section{Star Formation Rate Tracers}

At the most basic level, star formation rate (SFR) indicators are merely measurements of luminosity, either mono--chromatic or integrated over some specific wavelength range. The main target is to identify emission 
that probes recent star formation, while avoiding as much as possible contributions from more evolved 
stellar populations. This is generally accomplished by targeting continuum or line emission that 
is sensitive to the short--lived massive stars. 

Luminosities at all wavelengths across the 
electromagnetic spectrum, from the X--ray to the radio, have been employed to calibrate SFR 
indicators, targeting both the direct stellar emission in the UV/optical range and the dust--reprocessed 
stellar light in the mid/far--infrared  \citep[][and references therein]{Kennicutt1998b, Calzetti2009}. 
Because of observational limitations, SFR indicators at any wavelength have traditionally been reliably 
calibrated using the spatially--integrated light from luminous galaxies or luminous star--forming regions 
within galaxies. These calibrations have thus been luminosity--weighted towards the most active regions,  
also averaging across local variations in star formation history and physical conditions within each galaxy. 

The main limitation for both spatially--integrated and spatially--resolved SFR indicators is the 
presence of dust, which absorbs the light from stars. Furthermore, dust is more closely associated with 
star--forming regions, and there is a loose correlation between amount of dust extinction and star 
formation activity in both star--forming galaxies and regions \citep{Wang1996, Heckman1998, Calzetti2007}. Uniform infrared surveys such as the one provided by IRAS \citep{Soifer1986} have 
provided means to correct SFR indicators applied to whole galaxies. 

Until recently, spatially--resolved  measurements of SFRs had to rely on UV and/or optical tracers coupled to uncertain dust extinction corrections, due to the lack of high--angular resolution infrared measurements 
to probe the dust--obscured star formation. Over the past decade, however, the Infrared Space Observatory and the Spitzer Space Telescope  have transformed our approach to 
sub--galactic SFR measurements, by probing the dust--obscured star formation with a few arcsecond 
resolution, corresponding to $\lesssim$1~kpc size for galaxies within the Local Supercluster 
\citep{Kennicutt2003,Calzetti2005, Calzetti2007, Calzetti2009, Kennicutt2007, Kennicutt2009}.  
The soon--to--be--operational Herschel Space Telescope will 
expand on those capabilities, by spatially resolving  dust--obscured star formation at the peak energy 
emission ($\sim$70~$\mu$m--150~$\mu$m).

Corrections for the effects of dust, however, can be challenging when applied over sub--kpc regions. 
For instance, 
the stars that are responsible for the UV emission in a star--forming region are often spatially separated 
from the gas that emit in H$\alpha$ (and in the infrared), by a few tens to a few hundred pc \citep{Calzetti2005, Relano2009, Boquien2009}. Although this separation is irrelevant when 
measuring SFRs over entire galaxies, it becomes a crucial feature when the area over which the 
SFR is measured approaches the size of the star--forming region. In this case, the dust column density 
in front of the UV--emitting stars can be dramatically different from, and lower than, that in front of the  
H$\alpha$--emitting gas \citep{Relano2009}, by possibly a larger factor than what inferred from 
galaxy--integrated studies \citep[e.g.,][]{Calzetti1994}. 

SFR measurements of spatially--resolved regions within galaxies depend on many other physical factors besides dust attenuation. Because SFRs defined at different wavelengths 
probe different timescales (e.g., the UV continuum emission probes stellar populations in the age range 
$\approx$0--100~Myr, while the H$\alpha$ line emission probes the age range 
$\approx$0--10~Myr), factors such as local variations in the star formation history, star formation 
intensity, physical and chemical conditions, star cluster mass function, and stellar Initial Mass Function (IMF) are likely to play a role in the SFR calibrations. 

A variety of studies have recently established that traditional SFR(UV) and SFR(H$\alpha$) calibrations  
\citep[e.g.,][]{Kennicutt1998b}  yield different results when applied to bright or low--luminosity regions and galaxies. In bright galaxies SFR(UV)$\sim$SFR(H$\alpha$), implying that the underlying assumptions of those calibrations, i.e.  constant star formation over the stellar age range of interest 
and a universal stellar IMF \citep[e.g.,][or others]{Kroupa2001,Chabrier2003}, describe 
luminous galaxies reasonably well \citep[e.g.,][]{Salim2007,Meurer2009,Lee2009}.  Variations in SFR(UV)/SFR(H$\alpha$) due to differences in the adopted stellar population models are  around 
10\%--20\%, which is 
well within the scatter of the measurements in bright galaxies. However, during the past decade, observational evidence has been accumulating that, as their luminosity decreases, galaxies 
display a systematic trend for SFR(UV) to become larger than SFR(H$\alpha$) \citep[e.g.,][]{Sullivan2000,Bell2001,Salim2007,Meurer2009,Lee2009}. The discrepancy can be as large as an 
order of magnitude at the faintest end of the H$\alpha$ luminosity, as shown by  \citet{Lee2009} in 
an analysis of almost 350 galaxies within the local 11~Mpc\footnote{see, also, http://pompelmo.as.arizona.edu/~janice/11HUGS.html}. Preliminary analyses 
appear to indicate a similar trend between bright and faint sub--kpc regions {\em within} galaxies \citep{Boquien2009b}. Furthermore, in the last few years GALEX has discovered the 
existence of extended UV disks in nearby star--forming 
spirals, extended well beyond the ionized gas disk \citep{Thilker2005, Thilker2007, Dong2008}. 
 \citet{Salim2007} attribute the UV `excess' of low--luminosity galaxies to a luminosity--dependent excess attenuation correction in the UV data. More recently, \citet{Meurer2009} and \citet{Lee2009} have shown that the UV `excess' in faint galaxies is present {\em prior to dust attenuation corrections}.  

The existence of the problem is clear, but the determination of its nature will be considerably more 
difficult. Due to the different timescales they probe, the UV and H$\alpha$ emission are sensitive to the 
star formation history of the region; in the case of an instantaneous burst of star formation with 
fixed stellar IMF and stellar population model, by the time the H$\alpha$ intensity has decreased 
by two orders of magnitude, the UV has only decreased by a factor $\sim$6 \citep[using the 2007--updated models of][]{Leitherer1999}.  While the average star formation history 
of whole star--forming galaxies may be approximated by simple models like constant or exponentially 
declining star formation, the star formation history of small areas within those same galaxies is likely 
to be more stochastic in nature.  Nevertheless, star formation history may not  be the only 
necessary ingredient, as environment--dependent stellar IMFs are also a possible explanation to the 
observed effects \citep{Massey1995}.

If variations in the star formation history may account for the observed discrepancies between SFR(UV) 
and SFR(H$\alpha$) for resolved regions within galaxies, the same approach is less applicable 
for large samples of low--luminosity and/or low--surface--brightness galaxies, where such 
variations are expected to average out  or would imply implausible synchronizations among 
the galaxies \citep{Hoversten2008, Meurer2009}.  Alternative scenarios include a 
steepening of the high--end of the stellar IMF as a function of decreasing galaxy luminosity \citep{
Hoversten2008,Meurer2009}, and an environment--dependent  star cluster 
mass function, for which less massive galaxies do not form massive gas clouds, leading to an stochastic 
sampling of the high end of the stellar IMF \citep{Pflamm-Altenburg2009, Lee2009}.  The two 
scenarios, albeit physically distinct, yield very similar observational results in terms of integrated 
fluxes or colors. A discrimination among the two will require direct (via star counts) measurements of 
stellar IMFs over the full parameter space of galactic environments, as found within the local 
$\approx$10--15~Mpc. 

We are obviously in front of a severe limitation in our ability to apply standard calibrations of SFR 
indicators to sub--galactic regions. The challenge over the next few years will be to answer the following 
question: can we and how do we measure SFRs in spatially--resolved regions of galaxies?

\section{Gas Tracers}

The 21--cm line and CO emission are used to trace the neutral atomic and molecular gas (densities $\approx$300~cm$^{-3}$) components in galaxies, respectively.  Denser molecular gas phases, 
$\gtrsim$3$\times$10$^4$~cm$^{-3}$, have been recently probed using tracers like HCN \citep{Gao2004}. 

Surveys using existing facilities, like the VLA, WSRT, ACTA, CARMA, IRAM, Nobeyama, JCMT, 
etc. have, in recent times, produced or are producing homogeneous maps 
in HI and CO for nearby, luminous galaxies, in some cases with a few arcsecond resolution, to name a 
few, THINGS \citep{Walter2008}, BIMA--SONG \citep{Helfer2003}, HERACLES \citep{Leroy2009}, STING \citep{Rahman2009}. 

While the need for homogeneously observed, reduced, and calibrated maps of large samples 
of nearby galaxies is acute for both the atomic and molecular gas components, most of the 
challenges lay with the latter.  Even with today' s facilities and instruments, most CO maps trace the 
bright central regions and spiral arms of luminous galaxies. Conspicuously  absent, because generally 
undetected, are the interarm and outer regions of spiral galaxies and the dwarf and low surface 
brightness galaxies. 

Those missing portions of the parameter space are due to the combination of two factors: (1) lack 
of sufficient sensitivity with existing facilities; (2) the uncertain relation between CO line 
intensity and molecular hydrogen column density.  Maps of the nearby spiral NGC5194 obtained with 
the 45--m single--dish antenna of the Nobeyama Telescope \citep{Koda2009}  do indeed 
suggest that sensitivity to low--surface brightness emission is an important factor for detecting 
CO emission in faint galactic regions, possibly including the outer regions of large spirals. Observations 
with large single--antenna mm telescopes  \citep[e.g., the 
Large Millimeter Telescope,][]{Perez-Grovas2006, Schloerb2008} will be able to target low surface brightness 
emission in galaxies. In addition to the scaling laws, these maps will be instrumental 
for addressing the existence, universality,  and environmental and physical dependences of the 
threshold of star formation in galaxies \citep[e.g.,][]{Martin2001, Schaye2004, Boissier2007, Dong2008, Krumholz2008}. 

Far more complicated is determining whether a `universal' relation between H$_2$ 
column density and CO luminosity  (the X$_{CO}$--factor) is present in galaxies, and on which scales such relation would be applicable. Current determinations of X$_{CO}$ are mainly based on measurements 
made in luminous, metal--rich spirals and  range in value between 1.56~10$^{20}$ and 4.0~10$^{20}$~K~km~s$^{-1}$~cm$^2$ \citep{Bloemen1986, Strong1988, Young1991, Blitz2006, Draine2007}. One of the main caveats in the use of the X$_{CO}$ factor is its potential 
dependence on metallicity \citep[today still controversial, see,][]{Wilson1995, Boselli2002, Blitz2006}, 
and on the physical conditions of the molecular clouds \citep{Dickman1986}. One 
observational  result is that the detectability of CO decreases sharply with galaxy parameters loosely 
linked to luminosity, or mass, or surface brightness \citep{Meier2002, Leroy2009}. The 
self--shielding of CO is likely to decrease for decreasing metallicity, thus shrinking the physical size 
of the CO--emitting region in the molecular cloud. Studies of individual clouds in nearby galaxies 
covering the full parameter space of mass, luminosity, surface brightness, metallicity, etc. will be required to address these questions. 

Independently of how reliably the CO  traces H$_2$ under all or most 
conditions, searches for a complementary tracer of the molecular gas content of galaxies have 
become a timely endeavor. If the metal depletion on to dust is roughly constant from galaxy to galaxy, the expectation is that the dust--to--gas ratio will be proportional to the galaxy's or regions' s metallicity; 
this has been shown to be in reasonable agreement with the data, with a factor $\sim$2 dispersion,  
at least in a sample of nearby galaxies  \citep{Draine2007}. From that relation, the molecular gas 
content can be `reverse engineered', once metallicity, HI mass and dust mass are known.  To achieve 
this goal, accurate, sensitive, and high--angular resolution maps in both HI and dust emission will 
be required. Accurate determinations of dust masses on the scales relevant for probing the laws of star formation require observing, with $\sim$arcsec resolution, the full wavelength range 
from the infrared (starting around $\sim$10--30~$\mu$m) to the mm, where dust emission 
dominates over other processes. The Herschel Space Telescope only partially covers that requirement. 
The mm range is particularly important for measuring dust masses, since it probes the Rayleigh--Jeans 
tail of the black--body emission and is less sensitive to uncertain dust temperature(s) determinations 
\citep[e.g.,][]{Dunne2000}. 

The infrared/mm emission, however, only traces heated (by stars) dust, and should,
 technically, provide a lower limit to the actual dust content of a galaxy or a region. 
In addition to dust emission, molecular gas content can in principle be traced via dust absorption, 
which is related to the dust column density if the extinction law of a galaxy or a galactic region is 
known \citep[or reasonably determined,][]{Bohlin1978}. The main difficulties in applying this method 
to external galaxies are: isolating and measuring individual stars, both extincted and unextincted, 
and determining the line--of--sight location of those stars relative to the gas distribution. 

No less of an issue than measuring the spatially--resolved molecular gas content of galaxies is 
determining which gas density component  is most closely associated with the star formation. 
\citet{Gao2004} determined that in equation~1 the exponent N$\sim$1 if dense gas only, as 
traced by HCN, is considered. The nature of the gas component most closely associated with the star formation 
is still matter of intense debate, both theoretical and observational \citep[see, e.g.,][]{Blitz2006, Kennicutt2007, Narayanan2008, Leroy2008}. The biggest challenge 
for mapping the gas content of nearby galaxies and relating it to the physics of star formation remains 
securing uniform, high--spatial resolution, surveys of multiple gas tracers, probing different gas density 
phases, over a representative volume of the Local Universe.

\section{Summary}

New windows are being opened across the electromagnetic spectrum in the combined parameter 
space of  sensitivity and angular resolution by the refurbishment of the Hubble Space Telescope,  
the launch of the Herschel Space Telescope, and by the future space optical/infrared  (e.g., the James 
Webb Space Telescope) and ground millimeter and radio facilities (e.g., the Atacama Large Millimeter 
Array, the Large Millimeter Telescope, the EVLA, the Square Kilometre Array, etc.). These will enable  the 
investigation of  the accretion processes of neutral gas onto galaxies and of the physical mechanisms underlying the conversion of gas into stars, not only in our own Galaxy, but also  in nearby and distant galaxies as a function 
of cosmic time. 

With new opportunities come new challenges, both for star formation rate indicators and for gas tracers. 
As the focus of the field shifts from the analysis of galaxy--averaged quantities to spatially--resolved quantities within galaxies, calibrations of SFR indicators will have to be `adapted' for applications to 
sub--galactic regions. This will imply accounting for variations in dust column densities within resolved 
regions, as well as variations in physical and chemical conditions, star formation histories and, possibly, 
determining any environmental dependence of the stellar IMF and cluster mass function. 

For the gas tracers, challenges to be addressed over the next few years will include: determining whether the X$_{CO}$ factor is universal or is dependent on local conditions, and isolating 
which parameters it may depend on; testing alternative ways to trace the molecular gas content in 
galaxies; securing large surveys of nearby galaxies with uniform, sub--kpc resolution  maps 
covering the full parameter space of galaxy properties (luminosity, surface--brightness, mass, star 
formation intensity, global gas content, etc.), galactic conditions (including interarm regions of spirals, 
outer disk regions, etc.), and the full parameter space of gas conditions (density, metallicity, etc.) found 
in the Local Universe. 

Addressing the issues discussed in this short review will have far--reaching consequences for a 
number of fields investigating galaxies and galaxy populations. For instance, it will both provide the tools to 
interpret  observations of galaxies across cosmic times, from first light to the present, and input 
sub--galactic  star formation prescriptions for numerical and analytical simulations of galaxy formation 
and evolution. 

\acknowledgments

This work has been made possible by the efforts of two science teams: the SINGS (Spitzer Infrared 
Nearby Galaxies Survey) and the LVL (Local Volume Legacy) teams. SINGS and LVL are 
Spitzer Legacy programs; the Spitzer Space Telescope is operated by the Jet Propulsion Laboratory, 
California Institute of Technology under a contract with NASA.

D.C. thanks Ron Snell at the University of Massachusetts for many stimulating discussions on the 
relationship between CO and H$_2$.

\clearpage

\end{document}